\documentclass[aps,prl,twocolumn,showpacs,floatfix,preprintnumbers,amsmath,amssymb]{revtex4}
\usepackage{epsfig}
\usepackage{graphicx}
\usepackage{dcolumn}
\usepackage{bm}

\begin{document}

\title{1/f noise and multifractality in atmospheric-CO$_2$ records}
\author{Prabir K. Patra, M. S. Santhanam, P. Manimaran, M. Takigawa and T. Nakazawa}
\affiliation{Frontier Research Center for Global Change/JAMSTEC, Yokohama 236 0001, Japan,\\
Physical Research Laboratory, Navrangpura, Ahmedabad 308 009, India, \\
Centre for DNA Fingerprinting and Diagnostics, Hyderabad, India,\\
Center for Atmospheric and Oceanic Studies, Tohoku University, Sendai 980-8578, Japan}

\begin{abstract}
 We study the fluctuations in the measured atmospheric CO$_2$ records from
several stations and show that it displays $1/f$ noise and multifractality.
Using detrended fluctuation analysis and wavelet based methods, we  estimate the
scaling exponents at various time scales. We also simulate CO$_2$ time series
from an atmospheric chemistry-transport model (CTM) and show that eventhough
the model results are in
broad agreement with the measured exponents there are still some discrepancies
between them. The implications for sources
and sinks inversion of atmospheric-CO$_2$ is discussed.
\end{abstract}
\maketitle

  Atmospheric carbon dioxide is thought to be one of the greenhouse gases
primarily responsible for the current phase of global
warming \cite{co2}. In view of this significance for the global climate and
life on earth, considerable effort has gone into measuring CO$_2$ from
various platforms in the last several decades.
Direct record of atmospheric CO$_2$ is available since the 1950s and hourly
time series are constructed more recently at several
measurement stations \cite{co2ts,data}.
In the context of atmospheric systems, where the mean behaviour is reasonably
well known from such long term records, the fluctuations decide
the actual outcome. However, the fluctuation properties
of CO$_2$ time series have not been studied in sufficient detail.
Such a study would help identify the class of process, e.g,
correlated random walk, Levy flight etc., underlying the CO$_2$ biogeochemistry
and will provide a clue if it is a self-organised critical system.
Secondly, this would serve as a diagnostic tool to compare the outputs
from {\it ab initio} atmospheric CTMs as well as to constrain
the processes that could be represented in them. This could lead to
better understanding of the global CO$_2$ cycle.

Scale invariance and $1/f$ noise are considered to be the signatures
of complex systems \cite{complex}.
In recent years, several time series from physiological, economic, physical 
systems including those of atmospheric parameters have been
subjected to scaling and spectral analysis \cite{dfa1}.
If $x(t)$ be the given time series, then the
scaling can be expressed as, $x(\alpha t) \approx \alpha^H x(t)$,
where the symbol $\approx$ denotes statistical equality and the
the scaling exponent $H$ is also called the Hurst exponent.
In many of naturally occuring phenomena,
a suitable measure of their fluctuations $F(L)$
displays a power law as a function of length $L$ of the time series,
i.e, $F(L) \propto L^{H}$. Physically this implies that if $0 < H < 1$ then
they are long range correlated. In particular, $H=1/2$ corresponds
to white noise and $H=1$ to $1/f$ noise. A suitable stochastic model of long
range correlations is the fractional Brownian motion (fBm) \cite{dfa1}.
Significantly, many natural processes
such as the temperature and precipitation (as also many economic and
physiological time series) behave like a fBm process and exhibit long range
correlation with a exponent that is claimed to
be universal in some of those cases. For instance, in the case of
temperature fluctuations at inland
sites far from the oceans the exponent is $H \approx 0.65$ \cite{temp1}. See also
Ref \cite{temp2} for a debate. Similar
analysis of rainfall has been used to suggest that rain events
might be a case of self-organised criticality (SOC) \cite{rain}. Infact, the idea that
many natural processes might display $1/f$ noise and  SOC-like behaviour \cite{btw} is one
of the impetus for this line of research. However, to the best
of our knowledge, the fluctuation properties of the chemical species
in the atmosphere such as CO$_2$ have not been studied. In this letter,
we examine the experimentally measured high frequency (hourly) CO$_2$ time series
from 28 stations and show that their flucuations display a power law with
$H \approx 1$ corresponding to $f^{-\beta}$ ($\beta \approx 1$) noise at
intermediate frequencies. We also simultaneously examine the
model simulations of atmospheric-CO$_2$ and discuss its implications.

 The measured CO$_2$ concentrations constitute a non-stationary time series and hence
we apply the detrended fluctuation analysis (DFA) \cite{dfa2} to quantify the
fluctuations and its behaviour in the spectral domain.
An alternative approach is to use wavelets \cite{wlet} of appropriate basis
such that it allows for separation
between high and low frequency components through successive levels
of decomposition. We will only briefly mention the techniques here since
both the methods are widely reported in the literature \cite{dfa2,wlet}.
We denote by $x(k)$ the CO$_2$ concentration in parts per million (ppm)
measured at discrete times $k=1,2,....k_{max}$. The corresponding
integrated series given by, $y(t) = \sum_{k=1}^t x(k)$, $t=1,2....k_{max}$.
The detrending is done by dividing the series $y(t)$ into $M$ segments each
containing $\tau$ data points and by empirically fitting a linear, quadratic, cubic ...
function $\tilde{y}(t)$
to each of them. Then, the fluctuations about the fitted trend in each of the segment
is obtained as,
\begin{equation}
F(\tau) = \sqrt{ \frac{1}{\tau} \sum_{t=1}^{k_{max}} (y(t)- \tilde{y}(t))^2 }.
\end{equation}
The fluctuations $F(\tau)$
averaged over all the $M$ segments of size $\tau$ is denoted by $\langle F(\tau) \rangle$.
In many cases, $\langle F(\tau) \rangle \propto \tau^H$.
In the case of standard random walk, $H = 3/2$. Note that the
exponent $H$ is related to the power spectral exponent of $x(k)$ through
$\beta = 2H-1$ and to its fractal dimension $D=2-H$ \cite{dfa1}. We also obtain
the fluctuation function $F(\tau)$ using Daubechies wavelet \cite{wlet}, an
orthogonal wavelet family with largest number of vanishing wavelet coefficients.
We perform 9 levels of wavelet decomposition corresponding to quadratic
detrending.

\begin{figure}[t]
\includegraphics[width=8.0cm]{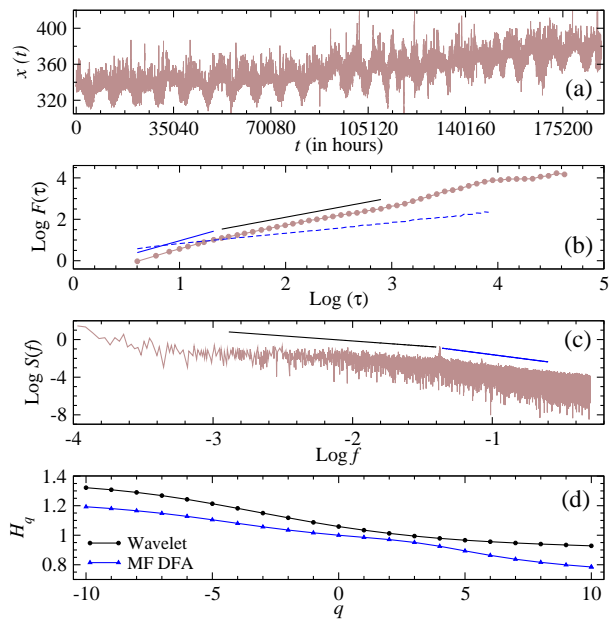}
\caption{(a) Hourly time series of measured CO$_2$ concentration at 
Brotjacklriegel, Germany. 
(b) DFA applied to this time series (solid dots) shows two distinct
time scales with different scaling exponents. The blue line is the 
best fit line with slope of 1.440 $\pm$ 0.012.
The black line fits the second region with slope of 0.953 $\pm$ 0.011. The
dashed line is the shuffled data and has a slope of 0.517, close to that of
white noise.
(c) The power spectrum of CO$_2$ series displays corresponding distinct 
frequency scaling. The blue line is obtained as the best fit one with
slope -1.920 $\pm$ 0.023
and the black line has the slope of -1.080 $\pm$ 0.032. The latter region
displays low-frequency $1/f$ noise. (d) Plot of $H_q$ against $q$ as a 
signature of multifractality.}
\label{ssl_dly}
\end{figure}

The hourly CO$_2$ data are available in the public domain for 28 stations,
some of which
have long period of measurements going as far back as 1972 \cite{data}.
As a typical case, Fig \ref{ssl_dly}(a) displays the time series of
CO$_2$ concentrations in ppm obtained at 
Brotjacklriegel, Germany. Apart from the hourly fluctuations, the profile
shows an approximate periodicity of about one year which is related to the
consumption pattern of CO$_2$ by the vegetation in the northern hemisphere.
In Fig \ref{ssl_dly}(b), we show the result of DFA performed on this data.
Two distinct scaling regimes can be identified. For a time scale of
few hours to two weeks, denoted by blue line in Fig \ref{ssl_dly}(b), we obtain
a best fit slope of 1.440 and this closely corresponds to 3/2 one gets for
a random walk process.
For a time scale of about 30 days (shown as black line),
the estimated slope is 0.953 and this regime closely corresponds to $1/f$ noise.
At longer time scales beyond $\tau > 300$ days,
we see an approximate white noise like behaviour with $H \approx 0.512$.
In order verify that these correlations are genuine, we create a surrogate
by randomly shuffling the series while retaining the same distribution. Random
shuffling would
kill the correlations present in the data. The DFA performed
on this shuffled surrogate data (shown as a dashed line) has a slope
of 0.517, in close agreement with the white noise exponent of 0.5.
We have also verified the DFA exponents using the wavelet decomposition
as well.
The power spectrum in Fig \ref{ssl_dly}(c) displays, as expected based on DFA
results,
$1/f$ noise at intermediate frequencies (black line) and $1/f^2$ behaviour at higher
frequencies (blue line). Notice that the estimated exponents $H$ in Fig \ref{ssl_dly}(b)
are related to the power spectral exponents $\beta$ in Fig \ref{ssl_dly}(c)
through $\beta = 2H-1$.
The results in Fig \ref{ssl_dly}(c) also indicate that at low frequencies
the spectrum is approximately flat representative of white noise. This succession of
three different
regimes in the power spectrum corresponding to white noise ($f^0$) at low,
1/f noise ($f^{-1}$) at intermediate
and random walk ($f^{-2}$) at high frequencies can
be generated from a superposition of exponential relaxation processes of
the type $N(t) = N_0 \exp(-\lambda t)$ with
the uniform distribution $g(\lambda)$ of relaxation rates $\lambda$ \cite{relax}.
While it might be tempting
to suggest this as a model for atmospheric-CO$_2$ variability, further work needs to
be done both at the level of statistics and much more in terms of
CO$_2$ science to understand the modeling aspect further.
Further, we will present results to show that this scaling behaviour
is generic to hourly CO$_2$ data from all the stations, with the
notable exception of Samoa and South Pole.

As shown above, Fig \ref{ssl_dly}(b) displays atleast three distinct
scaling regimes which indicates the possibility that CO$_2$ time series could be a
multifractal. In a monofractal, all the moments of the
fluctuations have the same scaling exponent $H$. For a multifractal, $q$th moment
is given by,
\begin{equation}
F(\tau) = \left( \frac{1}{\tau} \sum_{t=1}^{k_{max}} (y(t)- \tilde{y}(t))^q \right)^{1/q}
\propto \tau^{H_q}
\end{equation}
and the exponents $H_q$ are dependent on order $q$.
We compute $H_q$ using multifractal-DFA (MFDFA) and wavelet formalism \cite{mfdfa}
and in Fig \ref{ssl_dly}(d),
we show $H_q$ plotted against $q$. Note that at $q=2$, we have $H_2=1.062$, in
close agreement with $H$ obtained using DFA and wavelet methods earlier.
We will show that CO$_2$ series from all the stations display multifractal
characteristics.

\begin{figure}[t]
\includegraphics[width=7.5cm]{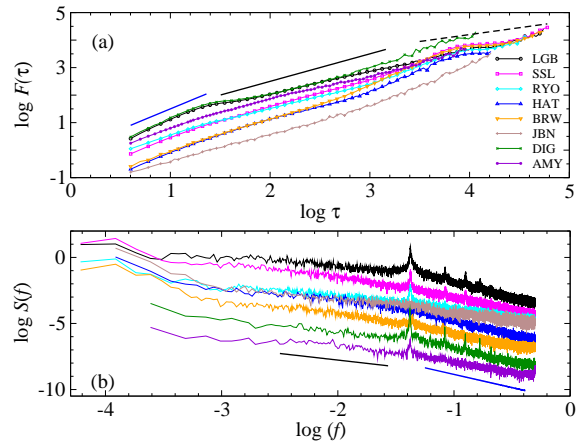}
\caption{(a) Fluctuation function $F(\tau)$ shown in log-log plot for seven
stations. Atleast two distinct scaling regions can be seen. Blue line has
a slope of 1.5 and the
black line has a slope of 1 meant for comparison purposes.
(b) Power spectrum for the same set of stations as in (a) in the same order in
which the station name appears in the legend. $y$-axis values are shifted for clarity.
The slope of black line is -1 and that of blue line is -2, both meant
for comparison with the power spectrum of actual data.}
\label{fig2}
\end{figure}

\begin{figure}[t]
\vspace{-0.5cm} \hspace*{-0.5cm}
\includegraphics[width=9.5cm]{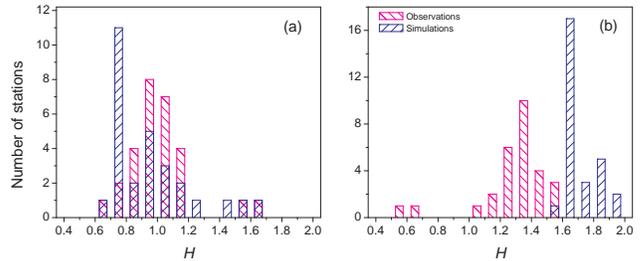}
\vspace{-1.2cm}
\caption{Histogram of DFA exponents for all the 28 stations, using both
model simulations (blue) and observations (magenta). (a) shows exponents
from $1/f$ region and (b) shows exponents from $1/f^2$ exponents 
(see text for details).}
\label{fig3}
\end{figure}

Our aim in rest of the study is to show that the scaling results
presented in Fig. \ref{ssl_dly} are generic to almost all the sites for which
the data is available. For this purpose, we consider the following
8 stations \cite{station} chosen to represent various geophysical features; namely,
Waldhof (LGB), Schauinsland (SSL) both in Germany, Ryori (RYO), Hateruma (HAT)
both in Japan, Barrow (BRW) in the USA,
Jubany (JBN) in Argentina, Puszeza Borecka (DIG) in Poland and Anmyeon-do (AMY) in
Republic of Korea. We show the DFA fluctuation functions
for these stations in Fig \ref{fig2}(a). In all the cases, atleast two distinct
scaling regimes are visible; the blue line represents the random
walk type scaling regime while the black line represents the
$1/f$ noise type scaling. The mean dfa exponent for these stations
in the former regime is $H \approx 1.404 \pm 0.125$ while for the
latter it is
$H \approx 0.978 \pm 0.087$ closely approximating unity necessary for
$1/f$ noise. The power spectrum shown in Fig \ref{fig2}(b) provides a
direct evidence of this. Beyond this, at longer time scales of about
130 days ($\log \tau > 3.5$),
some of the stations, AMY for instance, display white noise type scaling
as indicated by the dashed line. However, the white noise regime
does not seem to exist in all the stations.
In order to obtain a global picture, the histogram of DFA exponents
for all the stations is shown in Fig \ref{fig3}. Note that most of the
exponents assume a value between 0.8 and 1.1 which is our central result
that intermediate frequency CO$_2$ series exhibits $1/f$ noise. The power
spectrum of these stations (not shown here) also confirms the $1/f$ scaling.
In Fig \ref{fig3}(b), the histogram of $H$ for short time scales of less than
14 days ($1/f^2$ regime) is shown. Most of the exponents lie within the
range 1.3-1.7, in close vicinity of 1.5 needed for $1/f^2$ type power spectrum.

Thus, Figs \ref{fig2} and \ref{fig3} taken together reveal $1/f$ noise
(indicated as black line) for a time scale of approximately 30 - 50 days
in high frequency CO$_2$ records and it joins the host of other phenomena
that display $1/f$ behaviour \cite{complex}. Performing similar analysis on
the daily averaged data (not shown here) shows up $1/f$ noise on similar
time scales.

\begin{figure}[t]
\includegraphics[width=8.5cm]{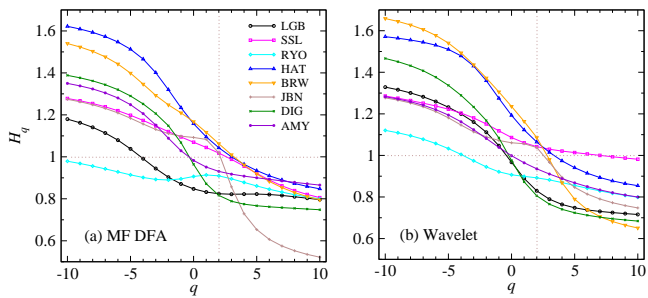}
\caption{Dimension spectrum as a signature of multifractality at intermediate
frequencies ($1/f$ regime at $q=2$) for the
same set of stations shown in
Fig \ref{fig2}. At $q=2$, most of the stations have $H_2 \approx 1$.}
\label{fig4}
\end{figure}

The theoretical modelling of CO$_2$ fluxes is presently
an important activity and it reflects the current
understanding of global CO$_2$ cycle \cite{model}. In order to compare our
observation based results against a theoretical benchmark, we
perform scaling analysis using DFA on the time series
obtained from atmospheric CTM.
We simulate the atmospheric-CO$_2$ time series using a
transport model and prescribed surface fluxes due to oceanic exchange
(monthly), terrestrial biosphere (3-hourly) and annual mean
fossil fuel emission \cite{model1}. The transport model
is based on the CCSR/NIES/FRCGC atmospheric general circulation model
(AGCM) nudgged with NCEP-FNL analysed wind vectors and temperature
\cite{model2}. The AGCM includes transport of chemical constituents 
due to advection, convection, and turbulent mixing processes.  
The model is run at
T106 ($\sim$1.125$\times$1.125) horizontal resolutions and 32 vertical
layers for the period 2002-2003 following a spin-up of 2 years.
The DFA exponents obtained from the CO$_2$ time series simulated by
the model at hourly time interval are shown in Fig. \ref{fig3}. 
Most of the model simulated exponents lie within the bins 0.7-1.2 (shown
as blue line) though about half of them assume values in the
range of 0.7 to 0.8.  The overlap between the
two histograms in Fig \ref{fig3}(a) shows an overall broad agreement
along with the differences in details that reflects partly the modelling
deficiencies. For instance, it is known that the correlation between the 
measured CO$_2$ fluxes and the simulated ones for seasonal time scales 
(corresponding partly to $1/f$ scaling regime) is 0.8 or greater \cite{model}.
The histograms for the $f^{-2}$ random walk type regime
in Fig \ref{fig3}(b) shows that even more discrepancies exist between 
measured and model results. This is because the synoptic scale 
($\sim$7-10 days) variations in CO$_2$ are not reproduced by the model 
simulations (correlation 0.5 or smaller) \cite{model}.
With this short period of simulation experiment, namely 2 years,
it is not possible to observe the white noise regime.

The results presented above are generic for most of the stations
except remote sites like Samoa and South Pole.
Significant deviations are observed mainly in the
$f^{-2}$ regime at remote observation sites, roughly in proportion to the
distance from the regions of fastest CO$_2$ exchange, namely
the forested lands and countries with greater human activity.
For instance, the large deviations between the scaling exponents from modelled
and observed time series are seen for the periods until 2, 7, 8 and 74 days
at Izana, Samoa, Mauna Loa and South Pole, respectively.

This comparison results suggest that the transport and
flux models are fairly realistic over the region of fast and large
CO$_2$ exchanges at the surface. But further studies are required
to understand the behaviour of CO$_2$ variations at the remote sites
like Samoa and South Pole. Both these stations display significant
deviations from the scaling results presented above.
We believe, such information can be used as a diagnostic
tool for testing the terrestrial and ocean ecosystem flux model results
as appropriate. In sources and sinks inversion of atmospheric-CO$_2$,
the forward model simulation lengths of basis functions and presubtracted
fluxes are dictated by the flux memory in CO$_2$ time series \cite{enting}.
Thus such scaling analysis would help in defining
the forward simulation lengths and the time scale of
future flux inversion studies.

Further, we show that multifractality is generic to CO$_2$ time series.
We calculate $H_q$ for the observed data at stations shown in Fig 
\ref{fig4} using MFDFA and  wavelet formalism. We stress that the results
from multifractal-DFA \cite{mfdfa} and wavelets are quantitatively similar.
The MFDFA results from modelled CO$_2$ time series also exhibit 
multifractality, though 
they do not agree quantitatively with the observation based results.
This once again suggests that the models are picking up broadly the 
correct behaviour in atmospheric-CO$_2$ but missing out the details.

We have studied the scaling properties of high-frequency atmospheric-CO$_2$ time series 
at 28 stations around the globe. At intermediate time scales of about a month,
CO$_2$ data exhibits $1/f$ noise. At shorter time scales of about a week, we get
$1/f^2$ noise. We also show that the CO$_2$ series is a multifractal implying
that manty different time scales are present in the system. We also examine the
{\it ab initio} atmospheric chemistry model outputs and show that the model results
reasonably reproduce the statistical properties of the measured series.
The implications
of these results in estimation and analysis of surface fluxes are
discussed.

\acknowledgments This work is possible due to high precision measurements
of atmospheric-CO$_2$ by several groups around the globe (see the full
list in WMO/GAW/WDCGG, 2006). Their relentless
effort is greatly appreciated. We thank Hajime Akimoto for supporting
this research.


\begin{thebibliography}{}

\bibitem{co2}
{\it Climate Change 2001: The Scientific Basis}, J. T. Houghton 
{\it et. al.} (eds), (Cambridge University Press, Cambridge, 2002).

\bibitem{co2ts} 
C. D. Keeling {\it et. al.}, Tellus {\bf 28}(6), 538 (1976).

\bibitem{data} Available for 28 stations at http://gaw.kishou.go.jp.
See also {\it Green house gases and other atmospheric gases}, volume IV,
WMO/GAW/WDCGG No. 30, Japan Meteorological Agency, Tokyo, Japan (2006).

\bibitem{complex} H. J. Jensen, {\it Self-organised criticality; Emergent
Complex Behaviour in Physical and Biological Systems}, (Cambridge University
Press, UK, 1998).

\bibitem{dfa1}
{\it Fractals and chaos in geology and geophysics}, D. L. Turcotte,
(Cambridge University Press, Cambridge, 1997);
{\it Science of Disasters : Climate disruptions, heart attacks and
market crashes}, edited by A. Bunde, J. Kropp and H.J. Schnellhuber,
(Springer, Berlin, 2002); S. Havlin {\it et. al.,} Physica A, {\bf 273},
46 (1999); M. Ausloos, Physica A {bf 336}, 93 (2004).

\bibitem{temp1} Eva Koscielny-Bunde {\it et. al.,} Phys. Rev. Lett. {\bf 81}
729 (1998); K. Freadrich and R. Blender, Phys. Rev. Lett. {\bf 90},
108501 (2003).

\bibitem{temp2} A. Bunde {\it et. al.,}, Phys. Rev. Lett. {\bf 92} 039801 (2004);
K. Freadrich {\it et. al.,}, Phys. Rev. Lett. {\bf 92} 039802 (2004).

\bibitem{rain} Ole Peters and K. Christensen,  Phys. Rev. E {\bf 66}, 036120 (2002);
Jan W. Kantelhardt {\it et. al.,} J. Geophys. Res. {\bf 111}, D01106 (2006).

\bibitem{btw} P. Bak, C. Tang and K. Wiesenfeld, Phys. Rev. Lett. {\bf 59},
381 (1987); P. Bak and K. Chen, Sci. Am. {\bf 79} 46 (1991).

\bibitem{dfa2} C-K. Peng {\it et. al.,}, Phys. Rev. E {\bf 49}, 1685 (1994);
Zhi Chen {\it et. al.,}, Phys. Rev. E {\bf 65}, 041107 (2002).

\bibitem{wlet} I. Daubechies, Ten Lectures on wavelets, SIAM, Philadelphia (1992);
P. Manimaran {\it et. al.,}, J. Phys. A {\bf 39}, L599 (2006).

\bibitem{relax} E. Milotti, Phys. Rev. E {\bf 72}, 056701 (2005); B. Kaulakys
{\it et. al.,}, Phys. Rev. E {\bf 71}, 051105 (2005).

\bibitem{mfdfa} J. W. Kantelhardt {\it et. al.,}, Physica A {\bf 316}, 87 (2002).

\bibitem{station} Apart from these 8 stations, the other 20 that form the basis
for this study are DEU, NGL, SCH, WST, ZGT
all in Germany; COI, DDR, KIS, MKW, MNM, TKY, YON in Japan; Mauna Loa, Samoa in USA;
Izana in Spain; FDT in Romania; PAL in Finland and
SNB in Austria and South Pole (see \cite{data} for details).

\bibitem{model} C. Geels {\it et. al.,} Tellus B {\bf 56}(1), 35 (2004); 
P. K. Patra {\it et. al.,} Atmos. Chem. Phys. Discuss. 
{\bf 6}, 6801 (2006); R. M. Law {\it et. al.,} in preparation.
 
\bibitem{model1} T. Takahashi {\it et. al.,} Deep-sea Res. II {\bf 49}, 1601 (2002);
S. C. Olsen and J. T. Randerson, J. Geophys. Res. {\bf 109}, D02301 (2004);
A. L. Brenkert, Carbon Dioxide Information Analysis Center, ORNL, Oak Ridge (2003).

\bibitem{model2} A. Numaguti, M. Takahashi, T. Nakajima, A Sumi,
CGER's Supercomput. Monogr. Rep. {\bf I-3}, Tsukuba (1997);
M. Takigawa {\it et. al.,} J. Geophys. Res. {\bf 110}, D21313 (2005).

\bibitem{enting}
I. G. Enting, {\it Inverse Problems in Atmospheric Constituent Transport},
(Cambridge Univ. Press, Cambridge, 2002).

\end{thebibliography}
\end{document}